\documentclass[12pt,amssymb,amsmath]{iopart}
\usepackage{latexsym}
\usepackage{amscd} 
\usepackage{amsfonts}
\usepackage[T1]{fontenc}
\usepackage[latin1]{inputenc}
\usepackage{graphics}
\def\be{\begin{equation}}
\def\ee{\end{equation}}
\def\ba{\begin{eqnarray}}
\def\ea{\end{eqnarray}}
\def\>{\rangle}
\def\<{\langle}
\def\<{\langle}
\def\>{\rangle}
\begin{document}

\title{Comment on ``note on the derivative of the hyperbolic cotangent''}

\author{ILki~Kim}

\address{Department of Physics, North Carolina Central University, Durham, NC 27707,
         U.S.A.}
\ead{hannibal.ikim@gmail.com}
%
%
%
\begin{abstract}
In a couple of articles (Ford G W and O'Connell R F 1996 {\em
Nature} {\bf 380} 113 and 2002 {\em J. Phys. A}: {\em Math. Gen.}
{\bf 35} 4183) it was argued that the standard result for the
derivative of the hyperbolic cotangent in the literature, $d \coth
y/dy = -\mbox{csch}^2 y$ is incomplete and the correct expression
should have an additional term proportional to the Dirac delta
function. The purpose of this paper is to demonstrate that this
claim is incorrect.
\end{abstract}
\maketitle
%
In \cite{FOR96} Ford and O'Connell published the formula
\begin{equation}\label{eq:ford1}
    \frac{d}{dy} \coth y\; =\; -\mbox{csch}^2 y\, +\, 2\,\delta(y)
\end{equation}
for $y \in {\mathbb R}$. Their argument showing that it is correct
can be summarized as follows. A function $f(y) = \coth y$,
satisfying $f(-y) = -f(y)$, is identically rewritten as
\begin{equation}\label{eq:0}
    f(y)\; =\; \textrm{sign}(y)\,\left\{1 + g(|y|)\right\}
\end{equation}
where an auxiliary function $g(y) = 2/(e^{2 y} - 1)$, and
$\textrm{sign}(y) = \pm 1$ depending on the sign of $y$. It was then
claimed that while the derivative of the second term,
$\textrm{sign}(y)\,g(|y|)$ equals $-\mbox{csch}^2 y$ for every
$y\neq 0$, the derivative of the first term, $\textrm{sign}(y)\cdot
1$ yields $2 \delta(y)$. Consequently, the derivative of $f(y)$
should contain the extra delta-function term. It is clear that this
logic is faulty. Indeed, we could repeat the same argument for an
arbitrary odd function $f(y)$ with a constant $c \ne 0$ in place of
factor 1 of the first term, which leads to the conclusion that
$df/dy$ would contain a term $2 c\,\delta(y)$. The error is, of
course, in disregarding the delta-function term, $-2 \delta(y)$
stemming from the derivative of the second term (for its detail,
refer to equation (48) and the discussion thereafter in
\cite{EST02}). In fact, by means of a partial fraction series for
the hyperbolic cotangent \cite{GRA00}
\begin{equation}\label{eq:kim1}
    \coth y\; =\; \sum_{k = -\infty}^{\infty}\, \frac{1}{y + i k
    \pi}\,,
\end{equation}
we can explicitly show that $d \coth y/dy = -\mbox{csch}^2 y$. Here,
we used \cite{GRA00}
\begin{equation}\label{eq:kim2}
    \csc^2 y\; =\; \frac{1}{\sin^2 y}\; =\; \sum_{k = -\infty}^{\infty}\,
\frac{1}{(y - k \pi)^2}
\end{equation}
with $\sinh y = -i \sin i y$.

The derivation of equation (\ref{eq:ford1}) was motivated by the
need to verify consistency between a quantum statistical mechanical
quantity and its classical counterpart to be identical to the
expression obtained from the quantum-mechanical quantity in the
limit of $\hbar \to 0$. However, equation (3) in \cite{FOR96} (and
also equation (2.11) in \cite{FOR88}) for the autocorrelation of the
quantum-mechanical random force $\hat{F}(t)$ in the case of constant
friction $\zeta$,
\begin{eqnarray}
    \hspace*{-1.8cm}&& \frac{1}{2} \left\<\hat{F}(t)\,\hat{F}(t')\, +\,
    \hat{F}(t')\,\hat{F}(t)\right\>\, =\, \frac{\zeta}{\pi} \int_0^{\infty}
    d\omega\,\hbar \omega\,\coth(\hbar \omega/2 k T)\,\cos[\omega (t - t')]\label{eq:kim3}\\
    \hspace*{-1.8cm}&=& k T \zeta\,\frac{d}{dt} \coth[\pi k T (t - t')/\hbar]\label{eq:kim4}
\end{eqnarray}
is actually incomplete while equation (4) in \cite{FOR96} (and also
equation (2.12) in \cite{FOR88}) for its classical counterpart is
correctly given as $\<F(t) F(t')\> = 2 k T \zeta \delta(t - t')$.
The integrand in equation (\ref{eq:kim3}) is an even function with
respect to the integration variable $\omega$. Hence, the integral
(understood in the distributional sense) may be performed from
$\omega = -\infty$ to $+\infty$ and can be evaluated in closed form
with the aid of identity (\ref{eq:kim1}). We then obtain an
additional term to (\ref{eq:kim4}) for the autocorrelation, arising
exactly from $k = 0$ in (\ref{eq:kim1}),
\begin{equation}\label{eq:kim5}
    \frac{2\,k T\,\zeta}{\pi}\,\int_{0}^{\infty}
    d\omega\,\cos \omega (t - t')\; =\; 2\,k T\,\zeta\,
    \delta(t-t')\,.
\end{equation}
This precisely corresponds to the classical autocorrelation while
equation (\ref{eq:kim4}) represents quantum fluctuations for the
autocorrelation about the classical mean value (\ref{eq:kim5}) so
that (\ref{eq:kim4}) vanishes in the limit of $\hbar \to 0$. It was,
however, argued in \cite{EST02} ({\em cf}. equations (44) and (45)
therein) that simply from equations (\ref{eq:ford1}) and
(\ref{eq:kim4}), $-\mbox{csch}^2 y$ can be interpreted as the
portion of $d \coth y/d y$ representing quantum fluctuations about
the classical mean value $2 \delta(y)$ such that
$\int_{-\infty}^{\infty} dy\,\mbox{csch}^2 y = 0$. This claim is
obviously incorrect since the additional term (\ref{eq:kim5}) was
just omitted (note also that $\mbox{csch}^2 y$ is an even function
anyway). The expression $d \coth y/d y$ has no direct physical
meaning whatsoever.

In \cite{FOR02} Ford and O'Connell gave some additional detail on
the ``derivation'' of formula (\ref{eq:ford1}). Motivated by the
observation that $\coth y$ increases by $+2$ as $y$ goes from
$-\infty$ to $+\infty$, they identified $\coth y$ to
$\mbox{Re}\left\{\coth(y + i \epsilon)\right\}$ with $\epsilon \to
0$ where
\begin{equation}\label{eq:coth_representaion_distribution1}
    \coth(y + i \epsilon)\; =\; \frac{\sinh 2 y}{\cosh 2 y - \cos 2\epsilon} -
    i \frac{\sin 2\epsilon}{\cosh 2 y - \cos 2\epsilon}\,,
\end{equation}
and argued that $\mbox{Re}\left\{\coth(y + i \epsilon)\right\}$ can
be understood as a distribution to yield the additional term $2
\delta(y)$ for $d \coth/d y$. However, it is incorrect. To show
this, we consider the identity obtained from the interplay between
the distributions and the theory of moments \cite{KAN04}
\begin{equation}\label{eq:g_delta_representation1}
    g(\lambda\,y)\; =\; \sum_{n=0}^{\infty} \frac{(-1)^n\,\mu_n\,\delta^{(n)}(y)}{n!\,\lambda^{n+1}}
\end{equation}
where $\mu_n = \int_{-\infty}^{\infty} dy\,g(y)\,y^n$ are the
moments of $g(y)$. For $g(y) = \mbox{Im}\left\{\coth(y + i
\epsilon)\right\}$, we have $\mu_{2m+1} = 0$ and
\begin{equation}\label{eq:moment1}
    \mu_{2m}\; =\; \frac{2\,(-1)^{m+1}\,\pi^{2m+1}}{2m+1}\,B_{2m+1}
\end{equation}
where $B_n$ are the Bernoulli numbers. Here we used \cite{GRA00}
\begin{equation}\label{eq:integral1}
    \int_0^{\infty} dx\,\frac{x^{2m}}{\cosh x - \cos 2 a \pi}\; =\;
    \frac{2\,(2m)!}{\sin 2 a \pi} \sum_{k=1}^{\infty} \frac{\sin 2 k a
    \pi}{k^{2m+1}}
\end{equation}
for $a \ne 1/2$, and
\begin{equation}\label{eq:bernoulli_representation1}
    \sum_{k=1}^{\infty} \frac{\sin 2 k a \pi}{k^{2n+1}}\; =\;
    \frac{(-1)^{n+1}\,(2 \pi)^{2n+1}}{2\,(2n+1)!}\,B_{2n+1}(a)
\end{equation}
where $B_n(a)$ are the Bernoulli polynomials with the property,
$B_n(0) = B_n$. From (\ref{eq:coth_representaion_distribution1}) and
(\ref{eq:g_delta_representation1}) we then obtain
\begin{equation}\label{eq:coth_representation_distribution2}
    \hspace*{-1.5cm}\coth(y + i\epsilon)\; =\; \mbox{Re}\left\{\coth(y + i
    \epsilon)\right\}\, +\, i \sum_{m=0}^{\infty}
    \frac{2\,(-1)^m\,\pi^{2m+1}\,B_{2m+1}}{(2m)!\,(2m+1)}\,\delta^{(2m)}(y)\,.
\end{equation}
The first term on the right hand side corresponds to the principal
part ${\mathcal P}\left\{\coth(y + i \epsilon)\right\}$ only, which
vanishes at $y=0$, while the second term consists of the
distributional contributions as for the well-known identity $1/(y
\pm i \epsilon) = {\mathcal P}(1/y) \mp i \pi \delta(y)$ where
${\mathcal P}(1/y) = y/(y^2 + \epsilon^2)$. As a result, $\coth y$
cannot be identified with $\mbox{Re}\{\coth(y + i \epsilon)\}$, nor
can $\mbox{Re}\{\coth(y + i \epsilon)\}$ be understood as a
distribution. Further, by using identity (\ref{eq:kim1}) with $y + i
\epsilon$ in place of $y \in {\mathbb R}$ we can get
\begin{equation}\label{eq:coth_x_i_epsilon}
    \coth (y + i \epsilon)\; =\; \sum_{k = -\infty \atop k \ne 0}^{\infty}\,
    \frac{1}{y + i k \pi}\, +\, \frac{{\mathcal{P}}}{y} - i \pi \delta(y)
\end{equation}
and thus $\coth y - {\mathcal P}\left\{\coth (y + i
\epsilon)\right\} = 1/y - {\mathcal P}(1/y)$. This demonstrates that
$\coth y$ and its derivative cannot be distributions either as none
of $1/y, {\mathcal P}(1/y), {\mathcal P}\left\{\coth (y + i
\epsilon)\right\}$ and their derivatives can. In addition, we see
from identity (\ref{eq:kim1}) that $\coth(\pm \infty) = \pm 1 \ne
0$, as compared to $1/y = 0$ at $y = \pm \infty$, stem simply from a
sum of all terms except for $k = 0$. Therefore, the increase of
$\coth y$ by $+2$ as $y$ goes from $-\infty$ to $\infty$ is not
ascribed to the additional delta-function term in (\ref{eq:ford1})
which should be removed. Actually, $L(y) := \coth y - 1/y$ is called
the Langevin function and plays an important role in quantum
statistical mechanics \cite{GRE97}.

In \cite{EST02} how singular functions can define distributions was
discussed by regularizing the singularities in the {\em standard}
way, namely, by either analytical continuation or extraction of the
Hadamard finite part. Special attention was paid to the distribution
$(\mbox{csch}^2 y)_{st}$ obtained from the standard definitions. It
was then shown that $d \coth y/d y = -(\mbox{csch}^2 y)_{st}$ with
no additional delta-function term. It was also argued that the
classical mean value (\ref{eq:kim5}) can be extracted from equation
(\ref{eq:kim4}) or $\lambda\,g(\lambda (t-t')) = -k T
\zeta\,\lambda\,(\mbox{csch}^2[\lambda (t-t')])_{st}$ with $\lambda
= \pi k T/\hbar$ in the limit of $\lambda \to \infty$. Using the
moment asymptotic expansion (\ref{eq:g_delta_representation1}),
$\lambda\,g(\lambda (t-t'))|_{\lambda \to \infty}$ exactly reduces
to the classical value $-k T \zeta\,\mu_0\,\delta(t-t')$ with $\mu_0
= \int_{-\infty}^{\infty} (\mbox{csch}^2 y - 1/y^2)\,dy = -2$ (refer
to equations (49)-(53) in \cite{EST02}). However, it may be argued,
from the physical viewpoint, that equation (\ref{eq:kim4}) at $t=t'$
cannot reflect the semiclassical limit $\hbar \to 0$, corresponding
to $t-t' \gg 1$, in which (\ref{eq:kim4}) identically vanishes. As a
result, this distributional concept is not needed to sustain the
consistency between classical and quantum autocorrelation in the
limit of $\hbar \to 0$ when the additional term (\ref{eq:kim5}) is
correctly considered.

The author is grateful to S. Winitzki (Munich) for stimulating
discussions.


\section*{References}
\begin{thebibliography}{10}
%
\bibitem{FOR96} Ford G W and O'Connell R F 1996 {\em Nature} {\bf 380} 113
%
\bibitem{EST02} Estrada R and Fulling S A 2002 {\em J. Phys.~A: Math.~Gen.} {\bf 35}
3079
%
\bibitem{GRA00} Gradshteyn I S and Ryzhik I M 2000
{\em Table of Integrals, Series, and Products} 6th edn (Academic
Press, San Diego)
%
\bibitem{FOR88} Ford G W, Lewis J T and O'Connell R F 1988 {\em Phys. Rev.} A {\bf 37} 4419-28
%
\bibitem{FOR02} Ford G W and O'Connell R F 2002 {\em J. Phys.~A: Math.~Gen.} {\bf 35} 4183
%
\bibitem{KAN04} Kanwal R P 2004 {\em Generalized Functions: Theory and
Applications} 3rd edn (Birku\"{a}user, Boston)
%
\bibitem{GRE97} Greiner W, Neise L and St\"{o}cker H 1997
{\em Thermodynamics and statistical mechanics} (Springer, New York)
%
\end{thebibliography}
\end{document}